\documentclass[conference]{IEEEtran}
\IEEEoverridecommandlockouts
\usepackage{cite}
\usepackage{amsmath,amssymb,amsfonts}
\usepackage{algorithmic}
\usepackage{graphicx}
\usepackage{textcomp}
\usepackage{xcolor}
\usepackage{booktabs}
\usepackage{multirow}
\def\BibTeX{{\rm B\kern-.05em{\sc i\kern-.025em b}\kern-.08em
    T\kern-.1667em\lower.7ex\hbox{E}\kern-.125emX}}
\begin{document}

\title{Electrocardiogram  Classification and Visual Diagnosis of Atrial Fibrillation  with DenseECG}

\author{\IEEEauthorblockN{1\textsuperscript{st} Dacheng Chen}

\and
\IEEEauthorblockN{2\textsuperscript{nd} Dan Li}

\and
\IEEEauthorblockN{3\textsuperscript{rd} Xiuqin Xu}

\and
\IEEEauthorblockN{4\textsuperscript{th} Ruizhi Yang}

\and
\IEEEauthorblockN{5\textsuperscript{th} See-Kiong Ng}

\thanks{1\textsuperscript{st} ad 2\textsuperscript{nd} authors at Institute of Data Science, National University of Singapore. Emails: \{idscd,idsld\}@nus.edu.sg 

3\textsuperscript{rd} author  at School of Computing, National University of Singapore. Email:xiuqin.xu@u.nus.edu

5\textsuperscript{th} author at School of Computing and Institute of Data Science, National University of Singapore. Email:seekiong@nus.edu.sg
}

}

\maketitle

\begin{abstract}
Atrial Fibrillation (AF) is a common cardiac arrhythmia affecting a large number of people around the world. If left undetected, it will develop into chronic disability or even early mortality. However, patients who have this problem can barely feel its presence, especially in its early stage. A non-invasive, automatic and effective detection method is therefore needed to help early detection so that medical intervention can be implemented in time to prevent its progression.  

Electrocardiogram (ECG), which records the electrical activities of the heart, has been widely used for detecting the presence of AF. However, due to the subtle patterns of AF, the performance of  detection models have largely  depended on complicated data pre-processing and expertly engineered features.  In our work, we developed DenseECG, an end-to-end model based on a 5 layers 1D densely connected convolutional neural network. We trained our model using the publicly available dataset from 2017 PhysioNet Computing in Cardiology(CinC) Challenge containing 8528 single-lead ECG recordings of short term heart rhythms (9-61s). Our trained model was able to outperform the other state-of-the-art AF detection models on this dataset without complicated data pre-processing and expert-supervised  feature engineering.

In addition, we address the need for explainability and traceability of machine learning models for medical diagnosis by plotting the {\it class activation map} (CAM) of an input ECG sequence with our trained AF detection model. CAM had been used to localize the discriminative regions of input images for image classification tasks. In this work, we  adopted it to  highlight the discriminative regions of the ECG data to visually unveil where our AF detection model was focusing on in classifying an input ECG sequence for AF.

Furthermore, the DenseECG model is applied to another more imbalanced MIT-BTH dataset. Our proposed model with class weight significantly outperforms the state-of-the-art models.

\end{abstract}

\begin{IEEEkeywords}
atrial fibrillation, convolutional neural networks, deep learning, healthcare
\end{IEEEkeywords}

\section{Introduction}
Atrial Fibrillation (AF) is currently identified as the most prevalent cardiac arrhythmia affecting 33.5 million people or 2.5\% to 3.2\% of global population\cite{AB}\cite{RKB}. The AF ``epidemic" is  escalating, especially among seniors, with about 5 million new cases being identified annually \cite{CHN}. If left undetected and untreated, the clinical consequences of AF can include cryptogenic stroke, ischemic stroke, heart failure, cognitive decline, dementia and early mortality. In fact, AF currently contributes one third of stroke which is the fifth leading cause of death in USA and a leading cause of severe chronic disability\cite{JWA}\cite{CHN}. It is estimated that 180 billion US Dollars will be spent yearly by 2020 on patients diagnosed with AF\cite{JXJE}.

Early detection of AF's presence in its early stage can allow timely introduction of effective medical intervention to decelerate or stop the progression of the disease. Unfortunately, the early detection of AF is not easy. As a result, about one-third of patients who have this kind of arrhythmia are unaware of its presence. Hence, it is also termed as silent atrial fibrillation (SAF) \cite{DK}.
  
\begin{figure*}[h]
	\centering
	\includegraphics[width=\textwidth]{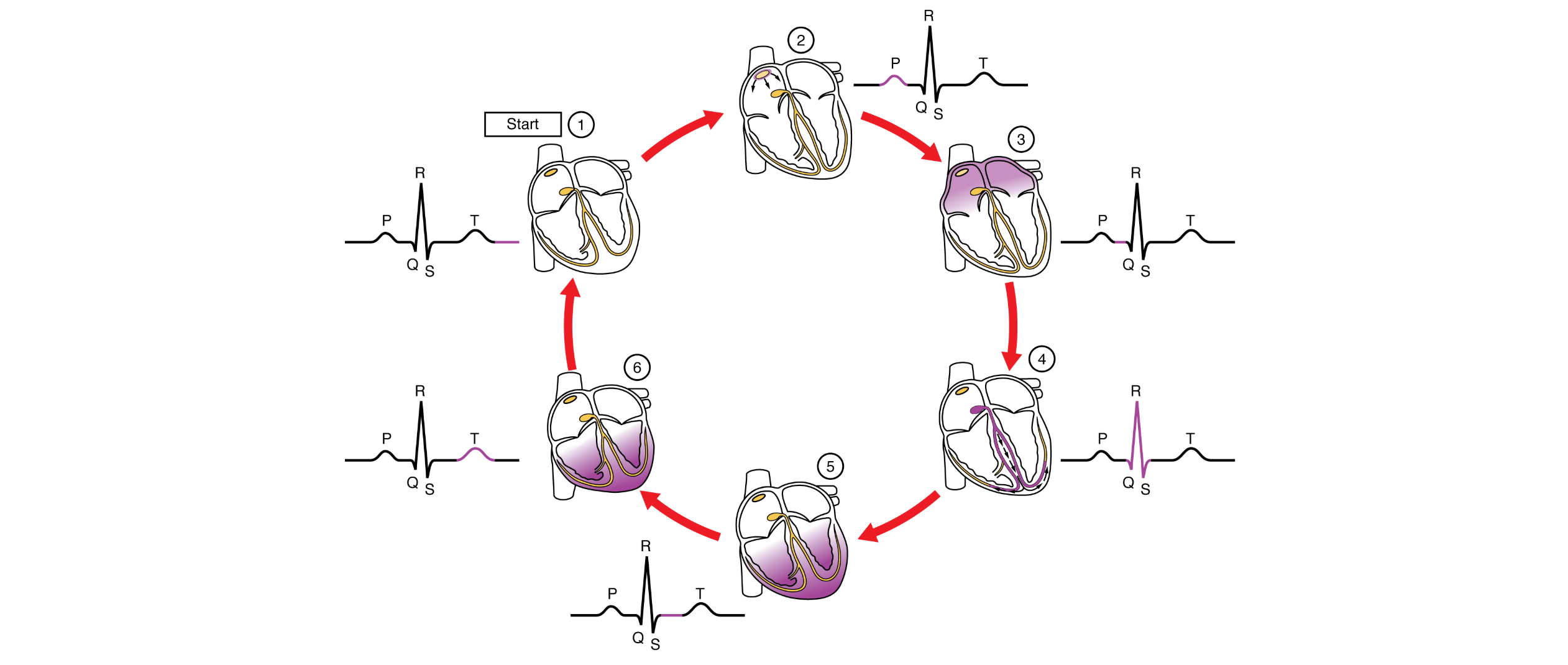}
	\caption{Normal Cardiac Cycle in association with ECG signal pattern\cite{Cardiac}\cite{BUGE}}
	\label{CARDI}
\end{figure*}

\begin{figure}[h]
	\centering
	\includegraphics[width=8cm]{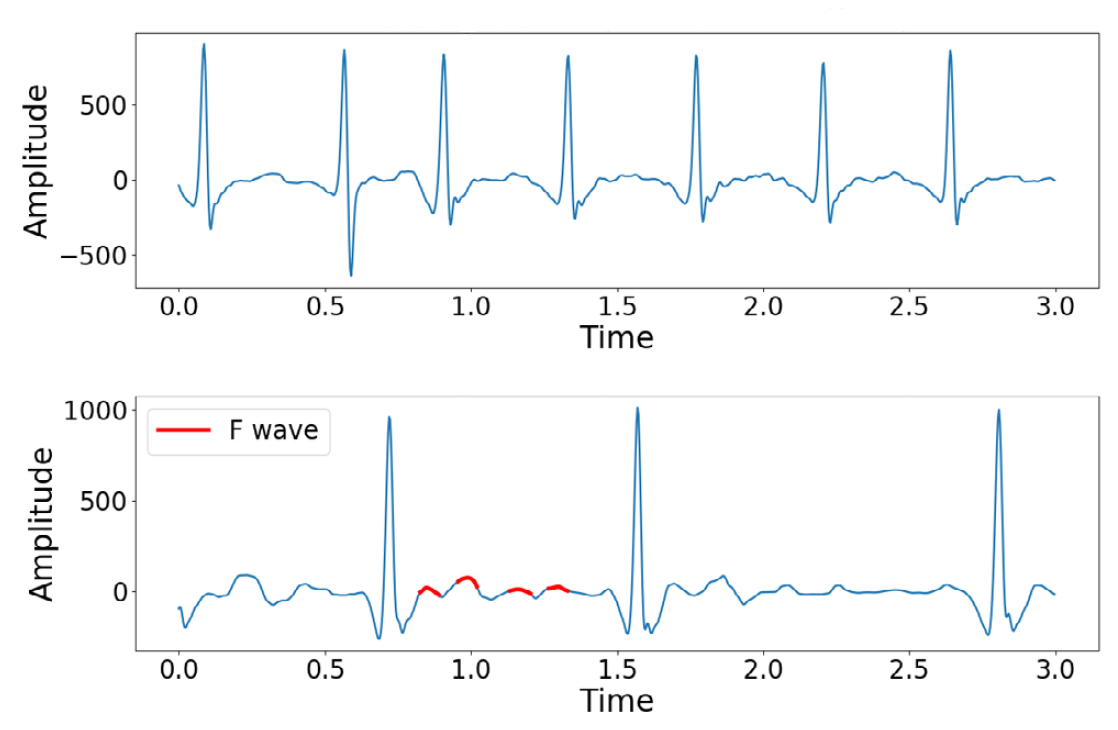}
	\caption{The above shows the absence of P wave and the below shows the presence of F wave \cite{CWJA} }
	\label{AFpic}
\end{figure}

Electrocardiograph (ECG),  which records the electrical activity of heart, is widely used for diagnosing AF.  Figure \ref{CARDI} shows the ECG signals with the actual heart contraction in a normal cardiac cycle. Under normal conditions, a cardiac cycle starts with a P wave (number 2 in the cycle, also called {\it atrial depolarization}) when sinoatrial (SA) node initiates the action potential that sweeps across the atria.  The P wave is followed by a small delay of activity (number 3 in the cycle) during which the atria is pumping blood. It is then followed by a QRS complex (also called {\it ventricular depolarization}, shown as number 4 in the cycle), and then followed by a small delay during which ventricles pumping blood, leading to a T wave (also called {\it ventricular repolarization}), during which ventricles relax. AF, which are described as `` {\it  irregularly irregular}'' beat rhythms, may be detected by looking for various tell-tale patterns or features  in the ECG recordings of the patients. Figure \ref{AFpic} shows some well-known example ECG patterns for diagnosing AF, namely  absence of P waves and appearance of F waves.

Such reading is clearly not easy for untrained eyes.  In fact, it can also be rather subjective, often resulting in different conclusions with multiple cardiologists. An automatic, accurate and less human-dependent way for diagnosing AF will save a significant amount of medical resources. However, due to the subtle patterns of AF, the performance of  detection models have largely  depended on complicated data pre-processing and expertly engineered features. For example,  hundreds of complex features  manually crafted by experts were required in order to train a classifier for the classification of AF\cite{TGCF}\cite{ZRKK}\cite{DPMB}. In addition, the trained models are often ``black boxes'', unable to make clear to the users the reasons with which a diagnosis is made by the computer models.  This is unacceptable for medical diagnosis in which the expanability and traceability of the model is of great importance if not mandatory.  As such, we seek to develop a machine learning model for AF detection that does not require  complicated data pre-processing and expert-supervised  feature engineering.  It should also be able to unveil which part of the ECG sequence that the AF detection model was focusing on in classifying an input ECG sequence for AF.

In this work, we have developed an end-to-end model for AF detection called  DenseECG based on a 5 layers 1D densely connected convolutional neural network.   We trained our model using the publicly available dataset from 2017 PhysioNet Computing in Cardiology(CinC) Challenge and showed that our trained model was able to outperform the other state-of-the-art AF detection models on this dataset.  The contributions of this work are two-fold:
\begin{enumerate}
	\item We have developed an end-to-end algorithm that can attain state-of-the-art performance for AF detection without depending on complex preprocessing or expert-supervised feature extraction;
	\item We have also adopted the {\it class activation map} or CAM plot of  DenseECG to reveal the discriminative regions of the input sequence.  We examined and compared the plot with actual medical practices that people use to identify AF, and showed that DenseECG was able to discover ECG regions from the data without requiring prior expert knowledge.
\end{enumerate}

\section{Background and Related works}
The study of AF can be traced all the way back to more than a century ago. The very first ECG showing AF was published by Willem Einthoven in 1906 \cite{EW} who invented the ECG and coined important related terms like P,Q,R,S,T waves. The link between irregularity of the pulse and AF was first discovered by Arthur Cushny\cite{AE} and Thomas Lewis \cite{LT}. 

For a long time, the diagnosis of AF was based on physicians' experience in reading  indicative features in ECG. It was only towards the end of last century that scientists began to develop quantifiable measures for the diagnosis of AF such as P-wave duration \cite{MJSG}\cite{NKPG}, P-wave dispersion \cite{MDMA}, P-morphology\cite{EPDZ}\cite{MZLP}\cite{ESVA}, Left ventricular hypertrophy (LHV) \cite{WTMC}\cite{CJSG} \cite{AKAS}\cite{MMSS} etc. However, as the ECG signals are often patient-dependent and highly variable in terms of scaling, displacement, noise, or even the patient's mood,   algorithms based on ad-hoc human-crafted features were not robust in detecting AF from ECG. As such, advanced signal processing methods such as wavelet transform, discrete Fourier etc were subsequently introduced for extracting features from ECG signals in the time and frequency domains \cite{AMM}\cite{MNT}\cite{KM}\cite{IGK} \cite{KAA}, as well as dimensionality reduction techniques for extracting features with higher discriminative power \cite{FW}\cite{OLM}. Typically, the extracted features were then used as inputs  to  discriminative algorithms such as support vector machine (SVM), hidden Markov models (HMM), etc to classify the ECG signals \cite{SKM}\cite{KPK} \cite{CR}\cite{CDR}.  

The advent of deep learning models  have led to great successes in  tasks including computer vision \cite{LSD}\cite{HZRS}, natural language processing \cite{SVL}\cite{CSWP} and time-series data analysis \cite{CPCS}\cite{YNSL}.   In recent years, researchers have also begun to develop deep models for AF classification.  While various types of neural networks with deep architectures (e.g. RBM, CNN, RNN) may be used for the classification of AF on the ECG data. The raw data are typically preprocessed and the proposed approaches mainly differ in the way the raw ECG data are preprocessed.    Some researchers preprocessed the raw data with `human experience', converting the raw data into hundreds of human-crafted features \cite{TGCF}.  Other researchers transformed the data with traditional signal processing techniques like wavelet transform, discrete Fourier transform, etc, converting the raw data into 2-D pictures before applying deep neural network classifiers \cite{PRRC}. Some  combining the above two approaches for the preprocessing \cite{HWZW}, instead of manually selecting and constructing the features, they used neural networks to automatically extract features. In \cite{XNCF}, authors input the data into a ResNet to do auto-feature extraction, then used RNN to do the classification. In this work, we seek to develop a deep learning model for AF detection without depending on such complex preprocessing or expert-supervised feature extraction. 

\section{Data}
Although the AF classification problem had been identified and studied by researchers for over a century, good-quality public datasets were hard to come by due to numerous reasons, including privacy concerns.    In this study, we use the dataset from PhysioNet/Computing in Cardiology Challenge 2017---a well-participated competition by many related researchers who used the more ``modern" methods like deep learning or deep learning hybrid with traditional methods.  

\begin{figure}[h]
	\centering
	\includegraphics[width=8cm]{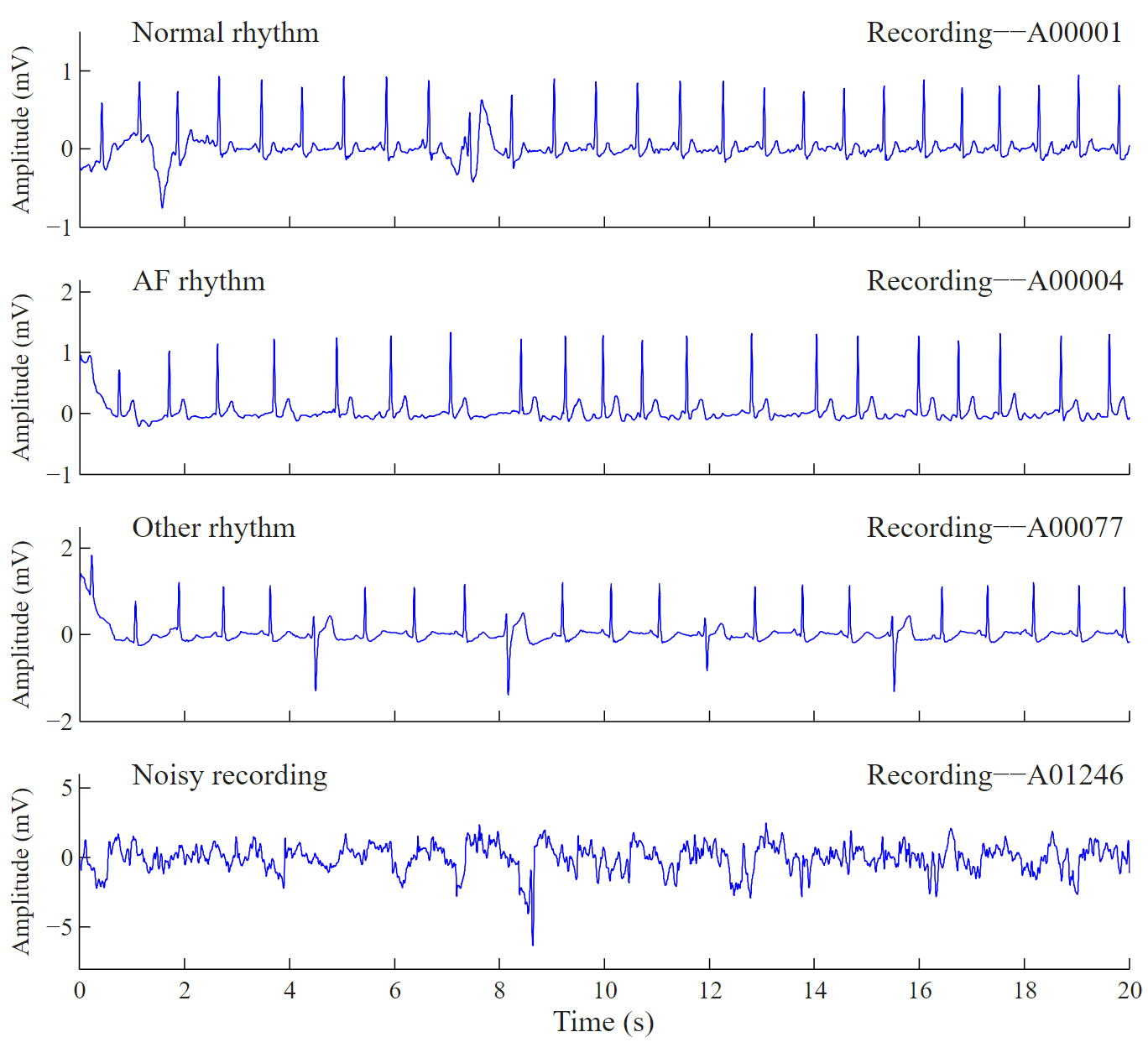}
	\caption{Plots of sample recordings from the four classes \cite{CWJA} }
	\label{ECGData}
\end{figure}

The ECG recordings in the dataset was collected using the AliveCor device, sampled at 300Hz. Each of the recordings was labeled by an expert using one of four labels: `Normal', `AF', `Other' or `Noisy', which were denoted by `N', `A', `O' and `P' respectively. The proportion of each class is shown in Table \ref{fourclas} and a visualization of the sample recordings is shown in Figure \ref{ECGData}.  In the aforementioned competition, a publicly available training set containing 8528 single-lead ECG recordings  lasting from 9s to 61s were given, while the test set comprising 3658 recordings of similar lengths(and class distribution) were kept private\cite{Cardiac}.
\begin{table}[h]
	\caption{Class partition of the ECG dataset}
	\begin{tabular}{cccccc}
		& Normal & AF    & Other  & Noisy & Total \\ \hline
		Count      & 5076   & 758   & 2415   & 279   & 8528  \\
		Proportion & 59.5\% & 8.9\% & 28.3\% & 3.3\% & 100\% \\ \hline
	\end{tabular}
	
	\label{fourclas}
\end{table}

\section{Method}

\subsection{Model Architecture}
Our proposed model, which we call DenseECG, follows the densely connected convolutional networks or DenseNets\cite{HLMW} that were primarily designed for image classification tasks and achieved state-of-the-art performance. In order to preserve information passing through multiple layers in a deep neural network such as ResNet \cite{HZRS} and Highway Networks \cite{SGS}, links that connect earlier layers to the later ones were added in a feed-forward fashion. DenseNets connects all layers directly with each other (as shown in the Figure \ref{dense} below) and is thus designed to mitigate the vanishing-gradient problem, strengthen feature propagation, encourage feature reuse and more importantly, enable the network to have very narrow layers, and hence also reduces the number of parameters\cite{HLMW}.

\begin{figure*}[]
	\centering
	\includegraphics[width=\textwidth]{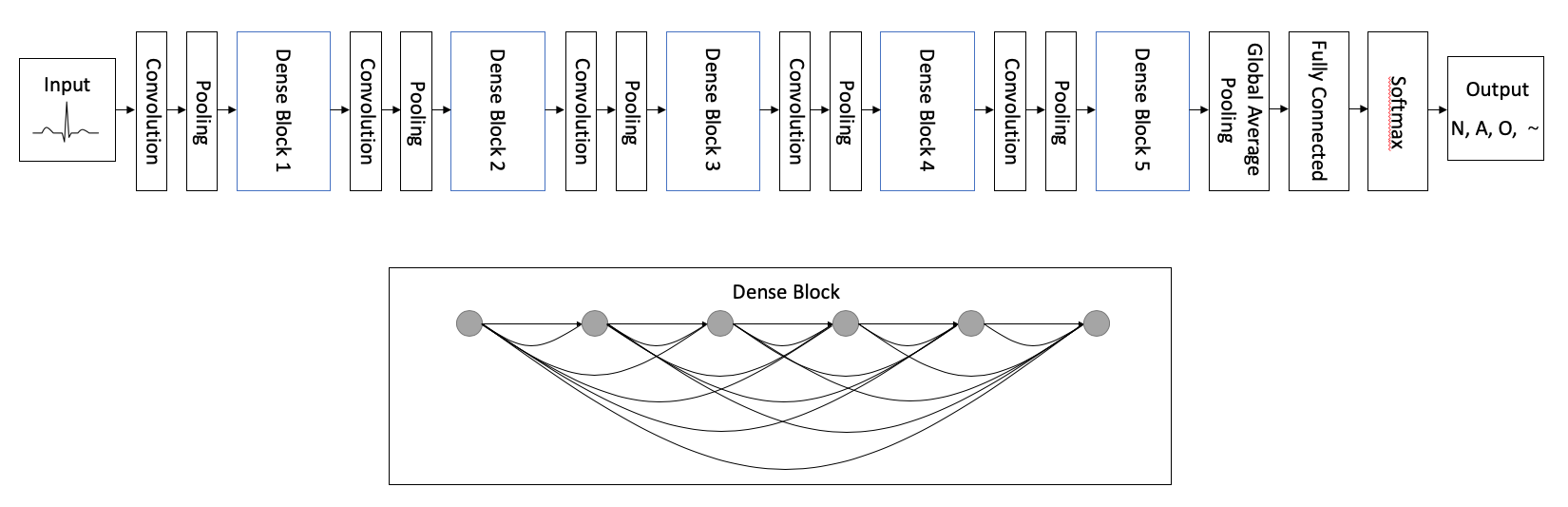}
	\caption{DenseECG model structure (above) and Dense Block structure(below)}
	\label{dense}
\end{figure*}

DenseNet in its original form has in fact been used by reseachers for  AF classification recently \cite{PRRC}.  The authors first preprocessed the ECG sequence through FFT into a 2D image, and then standard DenseNet was applied on it to do the classification.  In our work, we chose to exploit the structural advantage of DenseNet to avoid doing unnecessary data transformation which may incur potential information loss or introduce additional complexity. We adapted the structure to one dimensional input such that we can train it with   normalized time series directly. The structure of our network is  shown in Figure \ref{dense} (above).  There are a total of five dense blocks in our proposed structure.  Our choice was based on empirical observation through comparing structures with more or fewer blocks. Between the dense blocks, we have transition layers for convolution and pooling. At the end of the fifth dense block, the structure is followed by a standard set of global average pooling, fully connected layer and a softmax. 

Although structures like RNN, LSTM are usually used for sequence modeling, one advantage of CNN based models is that the class activation map can be easily obtained for visual analysis. This is  another reason for our model choice as it is important for medical diagnosis that we are able to highlight the discriminative regions of the ECG sequences that the model was using for its AF classification.

\subsection{Data Preprocessing}
As mentioned, unlike many current methods that required various filtering\cite{SSCN}, QRS detection\cite{CWJA}\cite{ZSR}, transformation\cite{PRRC}\cite{ZSR}, data augmentation\cite{PRRC}, etc, our objective is to avoid the dependence of complicated data pre-processing and expertly-engineered features for performance. As such, our data preparation and preprocessing is simple and minimal---we only normalized the data scales and padded all the data to the same length of 60 seconds (each recording is a one dimensional sequence of length 18000). The  data as processed were thereafter passed to the model for training and prediction.

\section{Results}
\subsection{Experimental Setup}
Given that the test dataset is not publicly accessible, researchers performed cross-validation on the training set, for example, \cite{XNCF}\cite{PRRC}\cite{XSZ}\cite{RPRC} used 5-fold cross validation, \cite{ZSR}\cite{CWJA} used 10-fold cross-validation. In this work, we performed both 5-fold and 10-fold cross validation to compare with the other works.  

In the training of our DenseECG model, Adam optimizer \cite{KB} was used and L2 regularizer was used for all convolutional layers with parameter $\lambda = 0.0001$. The learning rate was set at 0.001 and 25 epochs were trained. During the training process, we also used learning rate decay with a decay factor of 0.1. Decay happens twice at 50\% and 75\% of total epochs.

\subsection{Data Imbalance}
As shown in Table \ref{fourclas}, the dataset is not a balanced dataset---the number of Normal cases is much larger than the number of AF and Noisy. In one of the past works, \cite{VA} proposed to use oversampling technique to mitigate the data imbalance effect. Their oversampling was achieved by SMOTE (Synthetic Minority Over-Sampling Technique)\cite{CBHK}. They first generated a bigger data set, then trained on 85\% of the generated data set and tested on the remaining 15\%. Their model is a stack of three 1D CNN blocks and two layers of LSTM and have a performance of $F_1$ score 0.911, with $F_{1n}=0.922$, $F_{1a}=0.936$, $F_{1o}=0.802$. We also tried the similar practice with our DenseNet based model and found that it can drastically boost the result to a even higher level with $F_1$ score 0.922, $F_{1n}=0.940$, $F_{1a}=0.907$ and $F_{1o}=0.864$, which is even better than the result in \cite{VA}. However, when we did the data split prior to oversampling---for example, we first withheld 20\% of data, use the 80\% to do the oversampling and training, then we test the trained model on the withheld 20\% of data---no performance boosting can be seen anymore. Such phenomenon is in fact expected when oversampling was performed prior to data split---the performance boosting is well expected as the splitted testing portion will surely contain cases similar to the training set. However this is an over-estimation of the model's generalizability and the technique will not lead to any real improvement on the model's performance.

In this work, we use class weight to resolve the data imbalance problem. The class weights are calculated from the training samples: the smaller the size of one category, the larger class weight for the category. The class weights are then used to calculated a weighted loss of the samples during training so that the loss from smaller category such as `AF' has more importance than the larger category such as `Normal'. We have also tried focal loss \cite{lin2017focal} which reduces the relative loss for well-classfied example (in our case, the `Normal' category). The performance of focal loss does not help improve the performance for this dataset, thus we omit it and only report the results of DenseECG with and without class weight in the following.

\subsection{Performance and Comparisons}
First,  we use the performance measures as defined in the original competition \cite{Cardiac} to compare the various models that had been developed using the dataset.  The performance metric comprises  a set of $F_1$ values for each class and an average of $F_1$ of `Normal', `AF' and `Others'. Following the counting rules in Table \ref{count}, the $F_1$ values are defined as follows:

\begin{table}[h]
		\caption{Counting rules}
	\begin{tabular}{@{}ccccccc@{}}
		\toprule
		& \multicolumn{6}{c}{Predicted Class}                      \\ \midrule
		\multirow{6}{*}{\begin{tabular}[c]{@{}c@{}}Reference\\ \\ Class\end{tabular}} &        & Normal  & AF      & Other   & Noisy   & Total   \\ \cmidrule(l){2-7} 
		& Normal & Nn      & Na      & No      & Np      & $\sum$N \\
		& AF     & An      & Aa      & Ao      & Ap      & $\sum$A \\
		& Other  & On      & Oa      & Oo      & Op      & $\sum$O \\
		& Noisy  & Pn      & Pa      & Po      & Pp      & $\sum$P \\
		& Total  & $\sum$n & $\sum$a & $\sum$o & $\sum$p &         \\ \bottomrule
	\end{tabular}

	\label{count}
\end{table}
\begin{equation}
\begin{gathered}
{\text Normal{:} \; } F_{1n} = \frac{2\times Nn}{\sum N+\sum n},\;\; 
{\text AF{:} \; } F_{1a} = \frac{2\times Aa}{\sum A+\sum a}, \;\;\\
{\text Other{:} \; } F_{1o} = \frac{2\times Oo}{\sum O+\sum o}, \;\;
{\text Noisy{:} \; } F_{1p} = \frac{2\times Pp}{\sum P+\sum p}, \;\;\\
{\text Final\; score{:}\;} F_1 = \frac{F_{1n}+F_{1a}+F_{1o}}{3}
\end{gathered}
\end{equation}
where all the terms follow the naming rules in the Table \ref{count}.

In addition, we also report the average accuracy of the three classes `N',`A',`O', defined as follows: 
\begin{equation}
 \text{Average Accuracy} = \frac{\frac{N_n}{\sum N}+\frac{A_n}{\sum A}+\frac{On}{\sum O}}{3}
\end{equation}

\begin{table*}[]
\caption{Performance comparison with various previous works}
	\begin{tabular}{cccccccccc}
		\hline
		\multirow{2}{*}{Main Technique}                                                                                                                        & \multicolumn{5}{c}{Performance}                                                                                                                                                                                                                                   & 
		\multirow{2}{*}{CV}                                              
		& 
		\multirow{2}{*}{Data Set}                                         
		& 
		\multirow{2}{*}{\begin{tabular}[c]{@{}c@{}}Competition\\ Ranking\end{tabular}} 
		& 
		\multirow{2}{*}{Reference} \\
		& 
		$F_{1n}$                                                       & $F_{1a}$                                                       & $F_{1o}$                                                       & $F_1$                                                          &
		\begin{tabular}[c]{@{}c@{}}Average\\ Accuracy\end{tabular}      &                                                              &                                                               									   &      &                                                                                                           \\ \hline
		
		\begin{tabular}[c]{@{}c@{}}79 features crafted, classified with \\ XGBoost, RNN and LDA classifier\end{tabular}                                        & 0.903                                                          & 0.855                                                          & 0.736                                                          & 0.831                                                          &
		-& 
		-                                                              &     
		Test Set                                                       & 1                                                              & 
		\cite{TGCF}                     
		\\ \cline{1-1}
		
		\begin{tabular}[c]{@{}c@{}}491 features crafted, classified with\\ random forest\end{tabular}                                            & 
		\begin{tabular}[c]{@{}c@{}}0.909\\ (0.905)\end{tabular}        & 
		\begin{tabular}[c]{@{}c@{}}0.835\\ (0.794)\end{tabular}        & 
		\begin{tabular}[c]{@{}c@{}}0.734\\ (0.756)\end{tabular}        & 
		\begin{tabular}[c]{@{}c@{}}0.826\\ (0.818)\end{tabular}        &
		-&
		\begin{tabular}[c]{@{}c@{}}-\\ (10 fold)\end{tabular}             &   
		\begin{tabular}[c]{@{}c@{}}Test Set\\ (Training Set)\end{tabular} & 1                                                                 & 
		\cite{ZRKK}                   
		\\ \cline{1-1}
		\begin{tabular}[c]{@{}c@{}}150 features crafted, classified with \\ multilevel of AdaBoost\end{tabular}                                          &
		\begin{tabular}[c]{@{}c@{}}0.916\\ (0.909)\end{tabular}        & 
		\begin{tabular}[c]{@{}c@{}}0.823\\ (0.797)\end{tabular}        & 
		\begin{tabular}[c]{@{}c@{}}0.750\\ (0.772)\end{tabular}        & 
		\begin{tabular}[c]{@{}c@{}}0.829\\ (0.826)\end{tabular}        &
		-& 
		 \begin{tabular}[c]{@{}c@{}}-\\ (5 fold)\end{tabular}              &   \begin{tabular}[c]{@{}c@{}}Test Set\\ (Training Set)\end{tabular} & 1                                                                 & 
		 \cite{DPMB}                    
		 \\ \cline{1-1}
		 
		\begin{tabular}[c]{@{}c@{}}combine expert feature, centerwave feature\\  and deep feature(based on DNN), \\ classified with XGBoost\end{tabular}       & 0.912                                                          & 0.813                                                          & 0.751                                                          & 0.825                                                          & 
		-&
		-                                                                &  
		Test Set                                                         & 1                                                                & 
		\cite{HWZW}                     
		\\ \cline{1-1}
		
		Fine tuned CNN 13 layers                                                                                                                             & 0.920                                                        & 0.800                                                          & 0.790                                                          & 0.830                                                          &
		-& 
		5 fold                                                             & 
		Training Set                                                       & -                                                                  & \cite{KMA}                      
		\\ \cline{1-1}
		
		\begin{tabular}[c]{@{}c@{}}3 recurrent layers on top of \\ 16 residual blocks\end{tabular}                                            & 0.919                                                          & 0.858                                                          & 0.816                                                          & 0.864                                                          &
		-& 
		5 fold                                                            & 
		Training Set                                                      & -                                                                 & 
		\cite{XNCF}                     
		\\ \cline{1-1}
		
		\begin{tabular}[c]{@{}c@{}}188 features crafted, classified with\\ AdaBoost\end{tabular}                                          & 0.910                                                          & 0.860                                                          & 0.740                                                          & 0.825                                                          & 
		-&
		5 fold                                                             &
		Training Set                                                       & -                                                                  & \cite{MCDP}                     
		\\ \cline{1-1}
		
		\begin{tabular}[c]{@{}c@{}}external data augmentation, \\ ECG transformed to 2D spectrogram by FFT \\  trained and classify with DenseNet\end{tabular} & 0.910                                                          & 0.830                                                          & 0.720                                                          & 0.820                                                          &
		- &
		5 fold                                                            & 
		Training Set                                                      & -                                                                 & 
		\cite{PRRC}                     
		\\ \cline{1-1}
		
		\begin{tabular}[c]{@{}c@{}}30 features crafted, classified with\\ decision tree ensemble\end{tabular}                                          & 0.910                                                          & 0.820                                                          & 0.730                                                          & 0.820                                                          &
		- &
		100 fold                                                          & 
		Training Set                                                      & -                                                                 & 
		\cite{SBWB}                     
		\\ \cline{1-1}
		
		\begin{tabular}[c]{@{}c@{}}55 features crafted, classified with\\ SVM\end{tabular}                                               & 0.920                                                          & 0.820                                                          & 0.750                                                          & 0.830                                                          & 
		-&
		10 fold                                                           & 
		Training Set                                                      & -                                                                 & 
		\cite{BCLC}                     
		\\ \cline{1-1}
		
		\begin{tabular}[c]{@{}c@{}}ECG transformed into spectro-temporal\\ data matrix,  then trained and classify with \\ DenseNet\end{tabular}                         & 0.888                                                          & 0.796                                                          & 0.721                                                          & 0.802                                                          &
		- &
		10 fold                                                           & 
		Training Set                                                      & -                                                                 & 
		\cite{ZSR}                      
		\\ \hline
		
		\textbf{DenseECG, without class weight}                                                                                                                                     & 
		\textbf{\begin{tabular}[c]{@{}c@{}}0.929\\ 0.929\end{tabular}} & \textbf{\begin{tabular}[c]{@{}c@{}}0.863\\ 0.870\end{tabular}} & \textbf{\begin{tabular}[c]{@{}c@{}}0.825\\ 0.836\end{tabular}} & \textbf{\begin{tabular}[c]{@{}c@{}}0.872\\ 0.879\end{tabular}} &
		\textbf{\begin{tabular}[c]{@{}c@{}}0.872\\ 0.882\end{tabular}}&
		\textbf{\begin{tabular}[c]{@{}c@{}}5 fold\\ 10 fold\end{tabular}} & 
		\textbf{Training Set}                                             & -                                                                 & 
		\textbf{Our Work}          
		\\ \hline
		
		\textbf{DenseECG, with class weight}                                                                                                                                      & 
		\textbf{\begin{tabular}[c]{@{}c@{}}0.931\\0.931 \end{tabular}} & 
		\textbf{\begin{tabular}[c]{@{}c@{}}0.864\\ 0.870\end{tabular}} & 
		\textbf{\begin{tabular}[c]{@{}c@{}}0.821\\0.839 \end{tabular}} & 
		\textbf{\begin{tabular}[c]{@{}c@{}}0.872\\ 0.880\end{tabular}} &
		\textbf{\begin{tabular}[c]{@{}c@{}}0.885\\ 0.887\end{tabular}}& 
		\textbf{\begin{tabular}[c]{@{}c@{}}5 fold\\ 10 fold\end{tabular}} &  
		\textbf{Training Set}                                             & -                                                                 & 
		\textbf{Our Work}          
		\\ \hline
		
	\end{tabular}
	\label{performance}
\end{table*}

Table \ref{performance} lists the results reported by researchers on the dataset. We show the results of the 4 top ranking teams in the original competition in the first four rows of the table. After the competition, the dataset has continued to attract numerous groups of researchers working on  the problem of AF classification. We list the top performing ones (based on our knowledge, for the follow-up works were published in different venues) in the Table \ref{performance}.  As shown in Table \ref{performance}, the result of our proposed DenseECG shows that our model out-performed previous works\footnote{Note that only two of the four original winning teams reported their results based on cross-validation the training dataset.  While the other two wining teams did not provide such information, our performance based on cross-validation on the training set was better than the performance of the two teams reported on the unseen testing dataset.} in terms of $F_1$ scores.    It is important to note that many of the previous winning models required much in-depth domain knowledge for  data pre-processing and feature extraction.  Compared to all these methods, our model required the least expert input for feature extraction and minimal data pre-processing while achieving comparable or even better performance than all these reported methods. As such, manually-crafted feature with human experts' knowledge or complex data transformation  may not always be superior.  
The comparison of our model performance with the performance of \cite{PRRC} (row eight in Table \ref{performance}) is such a case in point. Here, the authors also used DenseNet and even used additional data. The complexity of their preprocessing was also much more sophisticated than ours: QRS detection, FFT versus to our simple normalization and padding. 

Comparing our two DenseECG models, it shows that the DenseECG with class weight has similar $F_1$ scores to that of the DenseECG without class weight, while the DenseECG with class weight has slightly higher average accuracy than the DenseECG without class weight. The improvement of average accuracy mainly comes from the `AF' category: an improving from 0.868 to 0.921 for the 5 fold cross-validation, and from 0.882 to 0.922 for the 10 fold classfication. In other words, the class weight helps improve the prediction accuracy for the categories with less samples, which is more important in practice.

\begin{figure}[!t]
	\centering
	\includegraphics[width=8cm]{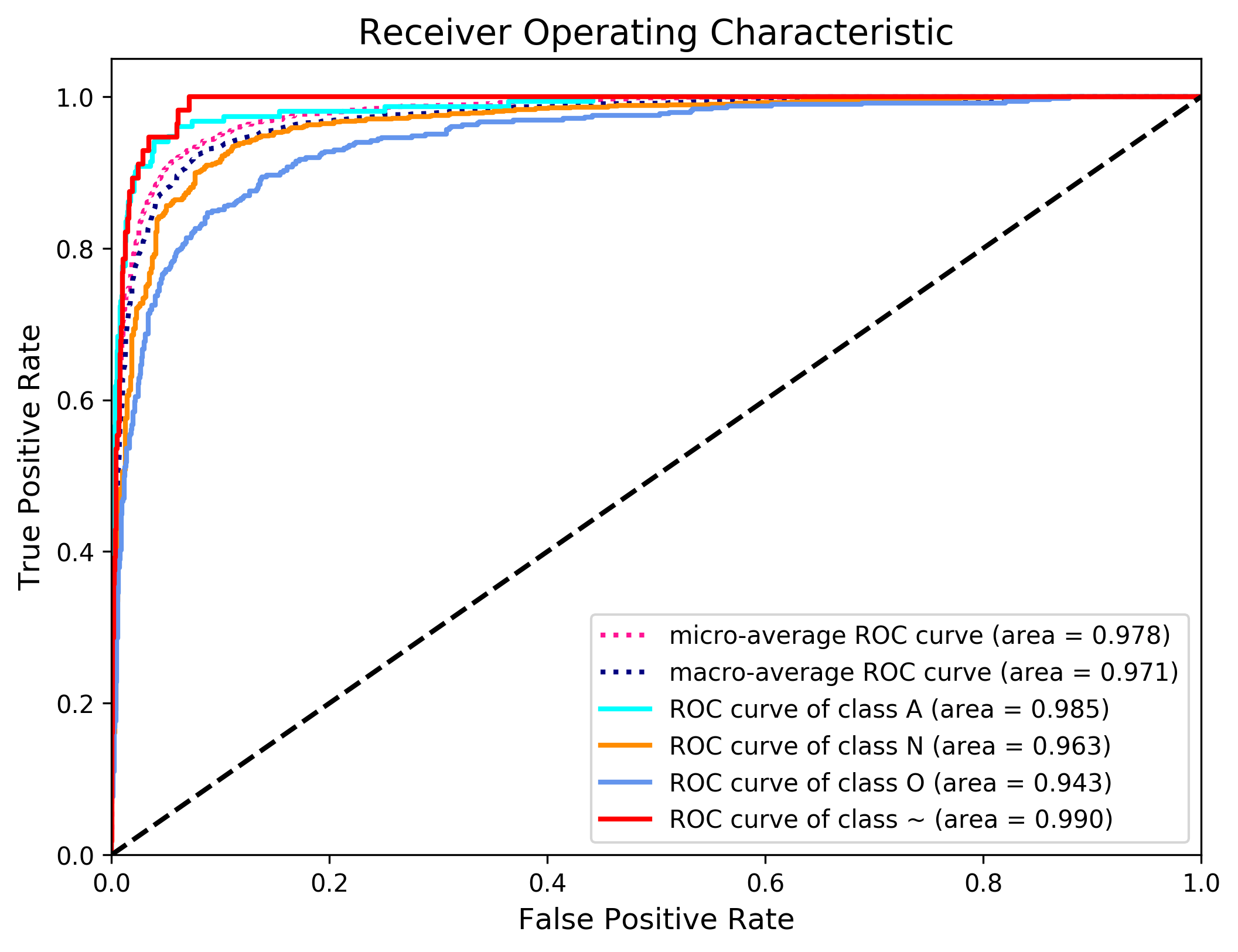}
	
	\caption{ROC curves for different classes and their averages}
	\label{roc}
\end{figure}

Apart from the $F_1$ measures used for the competition on this data set, some researchers have also reported the AUC of their methods on the dataset. As such, we also report the AUC of our trained model here\footnote{Five fold cross validation, the AUC is calculated based on the 20\% of data in the training set.}. The plot of Receiver Operating Characteristic (ROC) curves and their corresponding area under the curve (AUC) is shown in Figure \ref{roc}.

One of the latest research work for AF detection\cite{SSCN} used  24 hours Holter ECG data, which was much longer than the one minute data used in the competition.  Their reported AUC was 0.94 for testing set and 0.96 for training set (no cross validation was used,  the data were merely splitted into two). As shown in Figure \ref{roc}, In terms of AUC value, our DenseECG is higher. Even though given that the datasets were different, we cannot directly claim that DenseECG is better. Still, given that DenseECG had used much shorter ECG data sequence in a much noisier data set to achieve the reported AUC, we can reasonably expect DenseECG to be at least on par with if not better in terms of AUC if it were to be applied to the longer and cleaner ECG dataset.

\begin{figure*}[]
	\centering
	\includegraphics[width=\textwidth,height=2.1cm]{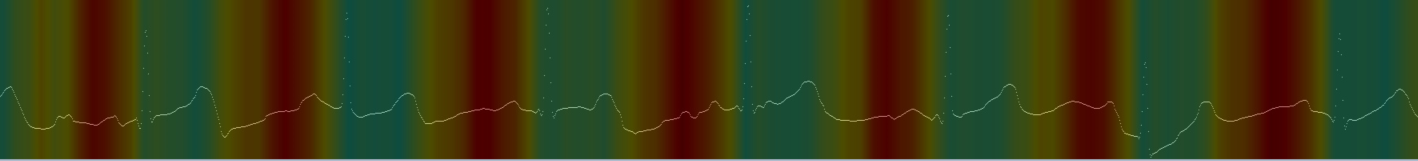}
	\includegraphics[width=\textwidth,height=2.2cm]{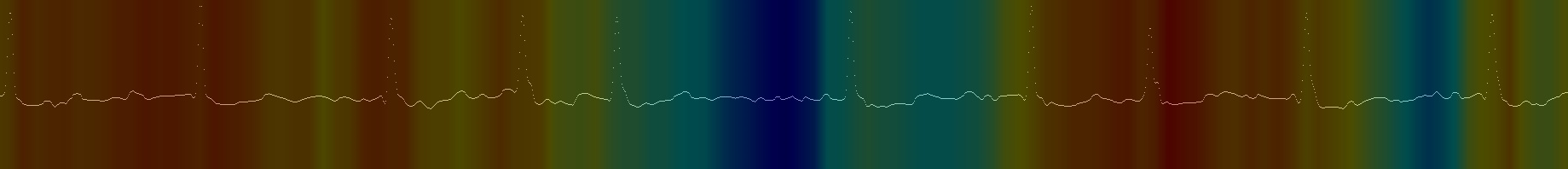}
	\includegraphics[width=\textwidth,height=2.2cm]{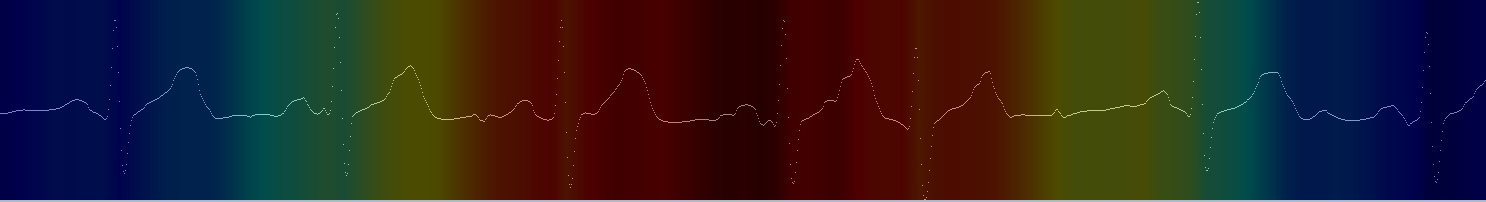}
	\includegraphics[width=\textwidth,height=2.2cm]{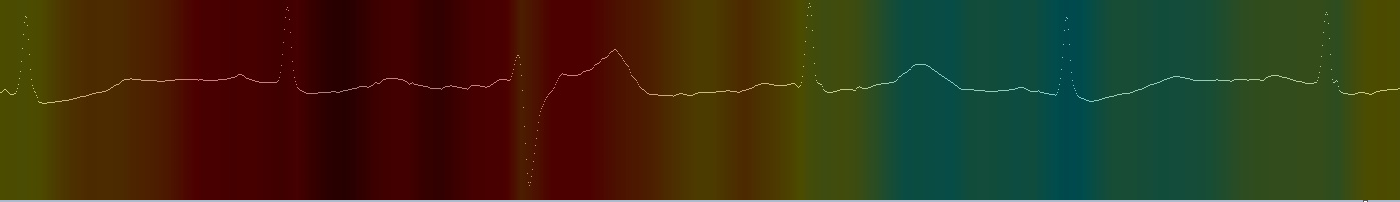}
	\caption{CAM for Normal (row 1), AF (row 2) and two samples from other (row 3 and 4)}
	\label{CAM}
\end{figure*}

\subsection{CAM Visualizations}
\subsubsection{Class Activation Maps}
A {\it global average pooling layer} that acts as a structural regularizer to prevent over-fitting is usually present at the latter stages of deep neural networks. The global average pooling layer can be used to derive a {\it class activation map} (CAM) \cite{ZKLO}  in image classification tasks (e.g. object detection in image scenes) to visualize the predicted class scores on a given input image, which provides insights on the actual regions within the input image where the classifier has focused its discriminative capacity on in order to make its classification decision.  In \cite{ZKLO}, the authors using CNNs trained on images with image-level labels (i.e. without any input on the object locations) to localize objects in image scenes, and showed that  the  CAM can  correctly highlight the discriminative regions of the input images.

In this work, we adapt the CAM approach to address the need for explainability and traceability of machine learning models for medical diagnosis.  Similar to the use of CAMs to localize the discriminative regions of input images for image classification tasks, we  can use it to shed light on how our AF detection model makes its decision by highlighting the discriminative regions of the ECG signals and visually unveil where our DenseECG is focusing on when classifying an input ECG sequence for AF.

To do so, we make use of the output of the last dense block of DenseECG which is a tensor of dimension $({\text batch\;size}, L, C)$, where $L$ is the length of the sequence after multiple layers of convolution or `abstraction', which is usually much shorter than the original input, $C$ is the number of channels for each `time-step' on this sequence. Let us use $f_c(t)$ to denote the activation of unit $c$ in this last convolutional layer at `time step' $t$. The global average pooling after this layer is thus $\sum_{t=1}^Lf_c(t)$. 

The pooling result is thereafter fed into a fully connected layers whose output dimension is the number of class $K$. These $K$ outputs, which we denote as $S_k\;k\in\{1,...,K\}$, are  used to calculate the softmax for each class later. In other words, 

\begin{equation}
S_k = \sum_c^Cw_c^k\sum_{t=1}^Lf_c(t) = \sum_{t=1}^L\sum_c^Cw_c^kf_c(t)
\end{equation}
where $w_c^k$ is the weight that corresponds to class $k$  for unit $c$. The class activation map for class $k$ can then be defined as: 

\begin{equation}
M_k(t) = \sum_{c}^C w_c^kf_c(t)
\end{equation}
Since $L$ is usually smaller than the initial input length of the sequence, the CAM is linearly interpolated to  align with the initial input sequence's length.

\subsubsection{Visualizations and Discussions}
To illustrate the use of CAMs for understanding AF detection by our DenseECG, we plotted the CAMs for the Normal, AF, and Other categories. As shown in Figure \ref{CAM}, starting from top to bottom, the first and second shows the CAMs of one sample each from the `Normal' and `AF' categories, while the third and fourth show two samples from the `Other' category.  In these plots, the red regions shows the discriminative regions of the input ECG data in which the network had focused its discriminative attention to make its AF detection decision. 

We can observe that in the CAM for Normal ECG, red regions periodically distribute alone the sequence. On closer inspection, we can see the red regions are all before the QRS complex, around the area of P wave and a portion before it. This is aligned with our knowledge on AF---as we have seen in Figure \ref{AFpic}, the major characteristics for manual AF identification includes P-wave absence and F wave presence, which occur exactly in this area highlighted by the red color in DenseECG's CAM for a normal input ECG. 

In contrast, we observe that in the CAM for AF, no clear patterns can be seen. Such plot corresponds to medical domain experts' description: ``irregularly irregular beat rhythm''. It is also interesting to note that the highlighted regions are without P-waves.

Compared to the `Normal' and `AF' categories, the `Other' category is probably more challenging for AF detection due to its inherent heterogeneity.  Multiple `other' cardiac arrhythmia are all classified under the label `Other', including: Tachycardia, Bradycardia, Atrial flutter, ventricular fusion beats and extrasystoles etc. For illustration, we show the CAMs for two `Other' samples in the third and fourth rows of Figure \ref{AFpic}.  The third CAM shows the presence of extrasystoles which is highlighted by the red regions of the CAM, while the fourth CAM  shows  ventricular fusion beat, also correctly highlighted by the CAM's red regions.  This is illustrative of the impressive discriminative capacity of our DenseECG for AF detection, given that all these regions were automatically and correctly detected by  DenseECG without any prior expert input and complicated pre-processing of the ECG data, and in the presence high data imbalance between AF and normal classes and hetereogenous classes under the `Other' label.

\section{Method generalizability with another dataset of cardiac arrhythmia}
\begin{figure}[!t]
	\centering
	\includegraphics[width=1\linewidth]{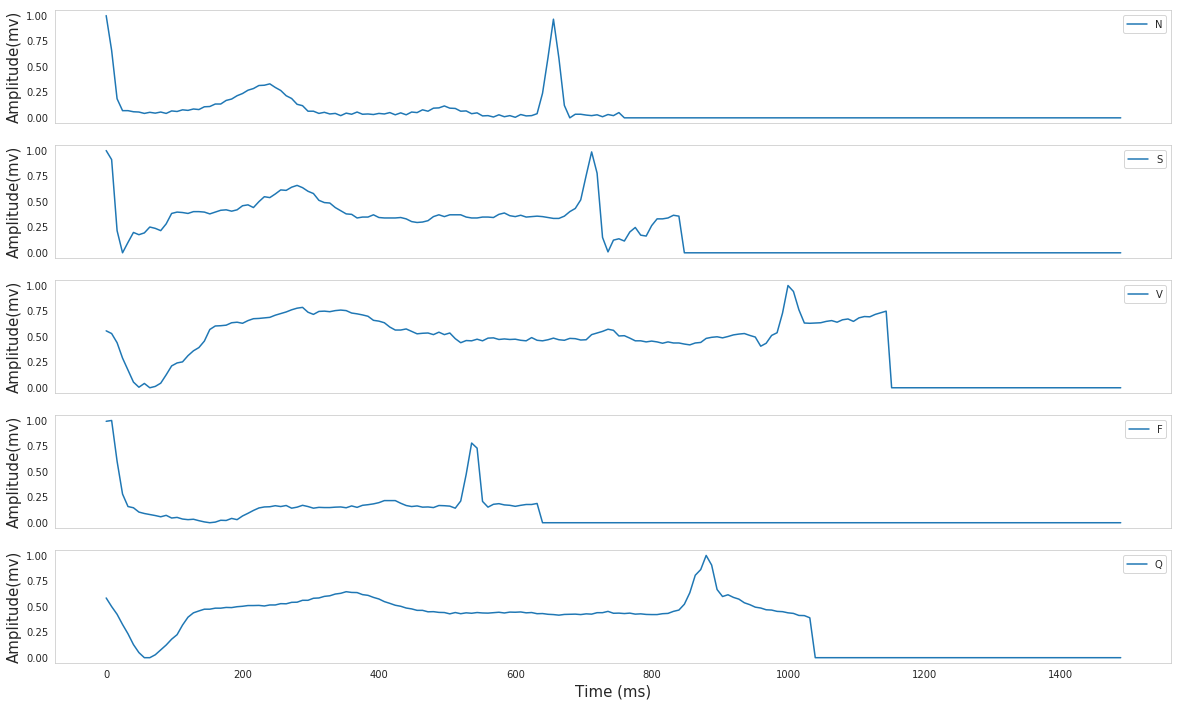}
	\caption{1-beat ECG for each category}
	\label{fig:mit-bih}
\end{figure}

Though our primary dataset is AF detection, it appears that the proposed DenseECG model can also be applied to  other cardiac arrhythmia classfication task. We further tested our method on another public dataset --  PhysioNet MIT-BIH Arrhythmia dataset, which has different kinds of cardiac arrhythmias. The MIT-BIH dataset consists of ECG recording of 47 subjects with a sampling rate of 360$Hz$ with annotations for each beat.
In order to have a fair comparison to \cite{kachuee2018ecg}, we directly used the processed data by \cite{kachuee2018ecg} which is available on \cite{MITBIH}. \cite{kachuee2018ecg} extracted each beat from the original data and using the original annotations to classify every beats to five categories, denoted by `N',`S',`V',`F',`Q', in accordance with Association for the Advancement of Medical Instrumentation(AAMI) EC57 standard \cite{association1998testing}. `N' stands for `Normal', while `S',`V',`F',`Q' are different kinds of cardiac arrhythmias, see Figure \ref{fig:mit-bih} for a visualization for each category. Each beat sample is resampled to the sampling frequency of 125 $Hz$ and padded to be with a fixed length of 187. The training and testing sample sizes are 87554 and 21892 respectively. The training samples are further stratified splited into training data and validation data with a ratio of 4:1. Data imbalance issue is more serious in the MIT-BIH dataset compared with the AF dataset: 82.2\% of the samples belong to `N', while only 0.7\%  of the samples are `F', see Table \ref{tab:fiveclas}.\cite{kachuee2018ecg} addressed the data imbalance via data augmentation. However, it is not clear which data augmentation technique was using and whether or not the augmentation technique would introduce bias into the results. We, on the other hand, impose class weights which are calculated from the training samples to each class to resolve the class imbalance issue. 

\begin{table}[!t]
	\caption{Class partition of the MIT-BIH dataset}
	\begin{tabular}{ccccccc}
		& N & S    & V  & F& Q&Total \\ \hline
		Count      & 90589   & 2779  & 7236   & 803   & 8039 & 109446\\
		Proportion & 82.8\% & 2.5\% & 6.6\% & 0.7\% & 7.3\% & 100\% \\ \hline
	\end{tabular}
	\label{tab:fiveclas}
\end{table}

As the length of each sample for the MIT-BIH dataset is 187, which is much smaller than 18000, the length of samples of the AF dataset, we use a much smaller DenseECG architecture for the MIT-BIH dataset compared with the AF dataset. The DenseECG architecture for MIT-BTH contains 3 dense blocks and each block consists of 3  CNN layers with a kernel size of 3, a growth rate of 12 and a dropout rate of 0.1. The reduction rate between dense blocks is chosen to be 0.25. The architecture is chosen based on the performance of the validation data. Adam optimizer \cite{KB} was used and L2 regularizer was used for all convolutional layers with parameter $\lambda = 0.0001$. The learning rate was set at 0.001 and 50 epochs were trained. During the training process, we also used learning rate decay with a decay factor of 0.1. Decay happens twice at 50\% and 75\% of total epochs. 

We use the average accuracy of the five classes with respect to the testing samples as the evaluation metric for the DenseECG model, in order to compared with previous works, see Table \ref{tab:mit_accuracy}. It shows that our proposed DenseECG model with class weight outperforms all the existing works, including \cite{kachuee2018ecg} which uses the ResNet on the same dataset but with a augmentation technique.
\begin{table}[htbp]
	\centering
	\caption{Comparision with previous works for the MIT-BIH dataset}
	\begin{tabular}{llr}
		\toprule
		\textbf{Work} & \textbf{Approach} & \multicolumn{1}{p{5em}}{\textbf{Average Accuracy (\%)}} \\
		\midrule
		Acharya et al \cite{acharya2017deep} & Augmentation + CNN & 93.50 \\
		Martis et al. \cite{martis2013application} & DWT + SVM & 93.80 \\
		Li et al. \cite{li2016ecg} & DWT + random forest & 94.60 \\
		Kachuee et al.\cite{kachuee2018ecg} & ResNet + Augmentation & 93.40 \\
		\textbf{Our work} & DenseNet + Class weight & \textbf{94.70} \\
		\bottomrule
	\end{tabular}%
	\label{tab:mit_accuracy}%
\end{table}%

In order to investigate the influence of different techniques used to address the data imbalance issue on the performance, we test different techniques including class weight and focal loss \cite{lin2017focal} on the MIT-BIH dataset with a ResNet and our proposed DenseNet respectively, see Table \ref{tab:mit_dataimbalance}. As the augmentation technique is not clearly elaborated in \cite{kachuee2018ecg},  we didn't try the augmentation with DenseNet and only report the result from \cite{kachuee2018ecg} with ResNet in the table. We use the same architecture and the same experiment set-up as described in \cite{kachuee2018ecg} for the ResNet. For the DenseNet, we use the architecture as described above. The ResNet and DenseNet have comparable number of parameters (50k). It shows that, both the focal loss and class weight can help improve the performance compared with models with no class weight, which overlooks the data imbalance problem. In addition, with the same technique, our DenseNet always outperforms its counterpart ResNet. Furthermore, class weight is better than the focal loss and the augmentation.
\begin{table}[htbp]
	\centering
	\caption{Comparison of average accuracy (\%) for different techniques to address data imbalance for the MIT-BIH dataset}
	\begin{tabular}{lrr}
		\toprule
		\textbf{Technique} & \multicolumn{1}{l}{\textbf{ResNet}} & \multicolumn{1}{l}{\textbf{DenseNet}} \\
		\midrule
		\textbf{No class weight} & 89.35 & 90.93 \\
		\textbf{Focal Loss} & 89.54 & 91.32 \\
		\textbf{Augmentation \cite{kachuee2018ecg}} & 93.40 & - \\
		\textbf{Class weight} & 94.47 & \textbf{94.70} \\
		\bottomrule
	\end{tabular}%
	\label{tab:mit_dataimbalance}%
\end{table}%

\section{Conclusions}
In this work, we have developed denseECG for the classification of ECG to detect AF. Our proposed model was able to achieve  state-of-the-art performance without expert-supervised feature engineering and sophisticated data pre-processing. In addition, we showed that CAM can be used for visualizing the regions in the input ECG sequence based on which that the model has made its decision, and that our DenseECG is able to correctly identify the regions that are important for AF detection, even for those complex ECG sequences classified under the  heterogeneous `Other' class.

Another advantage of DenseECG is that as shown in our experiment, it has worked well on rather short data lead,  which makes it  much easier to be incorporated into current systems where a sliding window of one minute can be easily obtained. We believe that long recordings (e.g. from Holter device) can similarly be processed so that the algorithm can  be implemented on wearable devices as well.  

In this study, while the class `Other' contained multiple cardiac arrhythmia other than AF, the corresponding CAM can still point to  meaningful problematic regions that  correspond to our human understanding of the disease. With datasets that contain all the explicit labels, more accurate class activation maps can be constructed to unveil more useful insights that will help in the diagnosis and even treatment of AF.

The DenseECG model is applied to the MIT-BTH dataset, a more imbalanced dataset with shorter data (around 1.5s per sample). Our proposed model with class weight is able to outperform the state-of-the-art models.

\end{document}